\documentclass[oldversion,preprint]{aa}  
\usepackage{graphicx}
\usepackage{txfonts}
\usepackage{longtable}
\usepackage{multirow}

\begin{document}

\title{VLBI observations of SN\,2011dh: imaging of the youngest radio supernova}

\titlerunning{VLBI observations of SN\,2011dh}

   \author{I. Mart\'i-Vidal\inst{1}
          \and
          V. Tudose\inst{2,3,4}
          \and
          Z. Paragi\inst{5}
          \and 
          J. Yang\inst{5}
          \and
          J. M. Marcaide\inst{6}
          \and
          J. C. Guirado\inst{6} 
          \and
          E. Ros\inst{6,1}
          \and
          A. Alberdi\inst{7}
          \and
          M. A. P\'erez-Torres\inst{7}
          \and
          M. K. Argo\inst{2}
          \and
          A. J. van der Horst\inst{8} 
          \and 
          M. A. Garrett\inst{2,9}
          \and
          C. J. Stockdale\inst{10}
          \and
          K. W. Weiler\inst{11}
}

   \institute{
             Max-Planck-Institut f\"ur Radioastronomie,
             Auf dem H\"ugel 69, D-53121 Bonn (Germany) 
             \email{imartiv@mpifr-bonn.mpg.de}
             \and
             Netherlands Institute for Radio Astronomy, 
             Oude Hoogeveensedijk 4, 7991 PD Dwingeloo (the Netherlands)
             \email{tudose@astron.nl}
             \and
             Astronomical Institute of the Romanian Academy, 
             Cutitul de Argint 5, RO-040557 Bucharest (Romania)
             \and
             Research Center for Atomic Physics and Astrophysics, 
             Atomistilor 405, RO-077125 Bucharest (Romania)
             \and
             Joint Institute for VLBI in Europe,
             Postbus 2, 7990 AA Dwingeloo (the Netherlands)
             \and
             Dpt. Astronomia i Astrof\'isica, Univ. Valencia,
             C/ Dr. Moliner 50, 46100 Burjassot (Spain)
             \and
             Instituto de Astrof\'isica de Andaluc\'ia, CSIC, 
             Apdo. Correos 3004, E-18080 Granada (Spain)
             \and
             Universities Space Research Association, NSSTC, 
             Huntsville, AL 35805 (USA)
             \and
             Leiden Observatory, Leiden University, 
             PO Box 9513, 2300RA Leiden (the Netherlands)
             \and
             Marquette University, Milwaukee, WI, USA
             \and
             Naval Research Laboratory, Washington D.C., USA
}

   \date{Letter accepted for publication in Astronomy \& Astrophysics}

  \abstract{
We report on the VLBI detection of supernova SN\,2011dh at 22\,GHz using a subset of the EVN array. 
The observations took place 14 days after the discovery of the supernova, thus resulting in a VLBI 
image of the youngest radio-loud supernova ever.
We provide revised coordinates for the supernova with milli-arcsecond precision, linked to the ICRF. 
The recovered flux density is a factor $\sim$2 below the EVLA flux density reported by other authors at 
the same frequency and epoch of our observations. This discrepancy could be due to extended 
emission detected with the EVLA or to calibration problems in the VLBI and/or EVLA observations.
}

   \maketitle

\section{Introduction}
\label{I}

Radio emission from core-collapse supernovae (CCSNe) is relatively rare. 
(Around 20--30\% of the CCSNe are detected in radio; see, e.g.,  Weiler et 
al. \cite{Weiler2002}.) There is, indeed, only a handful of these objects for 
which the radio structure has been (at least partially) resolved with 
very-long-baseline interferometry (VLBI) observations. 
However, it is definitely worth monitoring any potential radio emission from this
kind of events, in order to perform detailed studies of the physical conditions 
in the expanding supernova shocks.
The case of supernova SN\,1993J is the best example of such a study
(see, e.g., Bartel et al. \cite{Bartel2002}; Marcaide et al. \cite{Marcaide2010}; 
Mart\'i-Vidal et al. \cite{Marti2011a}; and references therein). The intense VLBI/VLA 
observing campaign of this supernova allowed these authors to constrain much of the parameter 
space of the models (density profiles of ejecta and circumstellar medium, hydrodynamical 
instabilities and their role in the magnetic-field amplification, 
energy equipartition, etc.) and to discover unexpected effects that led to the 
revision and extension of the standard supernova interaction model 
(Mart\'i-Vidal et al. \cite{Marti2011a}, \cite{Marti2011b}).

SN\,2011dh is a recent example of a radio-loud supernova. Located in the galaxy M\,51
(distance of 7--8\,Mpc; e.g., Tak\'ats \& Vink\'o \cite{Takats2006}) at the
coordinates $\alpha = 13^{\mathrm{h}}\,30^{\mathrm{m}}\,05.124^{\mathrm{s}}$ and
$\delta = +47^{\circ}\,10'\,11.301''$ (S\'arneczky et al. \cite{Sarneczky2011}), it 
was discovered with the Palomar Transient Factory project (PTF) on 2011 June 01 (Silverman 
et al. \cite{Silverman2011}; Arcavi et al. \cite{Arcavi2011}). 
Radio emission from SN\,2011dh was detected just three days after its discovery, with the 
Combined Array for Research in Millimeter-wave Astronomy (CARMA) 
at 86\,GHz (Horesh et al. \cite{Horesh2011}), and monitoring was also started with the 
Expanded Very Large Array (EVLA) and the Submillimeter Array (SMA)
at several frequencies, running from 8\,GHz to 107\,GHz (K.W. Weiler et al., in preparation).

The expansion velocity of the shock, as estimated from 
the H$\alpha$ blueshift, is 17\,600\,km\,s$^{-1}$ (Silverman et al. \cite{Silverman2011};
Arcavi et al. \cite{Arcavi2011}), 
similar to those of other supernovae of types II and Ib/c. Early X-ray emission was 
reported from {\em Swift} observations (Kasliwal \& Ofek \cite{Kaslival2011}). 
A good candidate for the progenitor star was isolated from {\em HST} 
observations (Li et al. \cite{Li2011}), and it shows very similar characteristics 
to the progenitor of the Type II-L (or spectroscopically-peculiar II-P) 
supernova SN\,2009kr (Elias-Rosa et al. \cite{Elias2010}). 

However, the results reported in Li et al. (\cite{Li2011}) on the progenitor of 
SN\,2011dh conflict with more recent results reported by Arcavi et al. (\cite{Arcavi2011}) and 
Soderberg et al. (\cite{Soderberg2011}). Based on a detailed monitoring of SN\,2011dh in X-rays 
(using the {\em Swift} and {\em Chandra} satellites) and radio (using the SMA, 
CARMA, and EVLA), Soderberg et al. (\cite{Soderberg2011}) report a fit to 
the data at all frequencies using a model of non-thermal synchrotron emission plus
inverse-Compton upscattering of a thermal population of optical photons. According
to these authors, SN\,2011dh would match a type IIb supernova better (i.e., similar 
to SN\,1993J), but with a compact progenitor and a shock expansion speed of 
$\sim$30\,000\,km\,s$^{-1}$ (a factor $\sim$2 higher than that of SN\,1993J).
Based on VLBI observations, Bietenholz et al. (\cite{Bieten2011}) do report
a barely resolved image of SN\,2011dh, on day 83 after the optical discovery, with 
an angular radius of 0.11$^{+0.09}_{-0.11}$\,mas (i.e., an average expansion speed of 
1.9$^{+1.6}_{-1.9} \times 10^4$\,km\,s$^{-1}$).

In this letter, we report on an earlier VLBI detection of SN\,2011dh made at the frequency of 
22\,GHz, just 14 days after the discovery of the supernova. We provide a revised position 
of the supernova with milli-arcsecond precision, based on phase-referencing observations 
with a calibrator in the international celestial reference frame (ICRF). This position may 
be useful for improving future VLBI observations of the supernova. (Indeed, the position estimate
reported here has been used in the correlation of the observations already reported in 
Bietenholz et al. \cite{Bieten2011}.) In the next section, we 
describe our observations and the calibration strategy followed in the data analysis. In 
Sect. \ref{III}, we report on the results obtained. In Sect. \ref{IV}, we summarize our 
conclusions.

\section{Observations and data reduction}
\label{II}

The observations were performed on 2011 June 14 using part of the European VLBI 
Network (EVN) at 22\,GHz. The participating antennas were Effelsberg (100\,m 
diameter, Germany), Robledo and Yebes (70\,m and 40\,m, respectively, Spain), Onsala 
(20\,m, Sweden), Metsah\"ovi (14\,m, Finland), and the MkII telescope at Jodrell Bank 
(25\,m, United Kingdom).

The recording rate was set to 1\,Gbps (dual-polarization mode), with a total bandwidth 
coverage of 128\,MHz (divided into eight equal sub-bands for the data recording) and a two-bit 
sampling. The observations lasted 11\,hours and 24\,minutes (about 120 hours of overall 
baseline time), but only a total of $\sim$70\,hours of useful baseline time was obtained 
after the data correlation and calibration, mainly due to the more limited participation 
of some of the stations and the nondetection of fringes related to Robledo (likely related
to issues in the antenna subreflector) and 
Metsah\"ovi (problem with MkIV formatter).

The observations were scheduled in phase-reference mode, using the source J1332+4722 
(about 0.5\,degrees away from the supernova) as the main phase calibrator. Additional sources
were observed as fringe finders and flux calibrators (3C286, 3C345, and J1156+295,
observed every 40--50 duty cycles), as a secondary calibrator (B1333+459, observed 
once every six to ten cycles), and to allow the calibration of the evolution in the tropospheric 
delay at each station (see, e.g., Brunthaler et al. \cite{Brunthaler2005} for details). Duty 
cycles with two different periods (90 and 120\,seconds) were used in the observations, 
since the coherence time and the optimum on-calibrator integration time for successful detections 
were not known with precision. All sources were also observed in single-dish mode at the Effelsberg 
radio telescope to have simultaneous estimates of the total flux densities of the calibrators.

The data were correlated at the Joint Institute for VLBI in Europe (JIVE, the Netherlands) 
using 128 channels per sub-band and an integration time of one second per visibility. Only the 
parallel hand data were correlated (i.e., the LCP and RCP data), but not their cross 
correlations. 

The data 
calibration was performed using the astronomical image processing system (AIPS) of the 
National Radio Astronomy Observatory (NRAO, USA) with standard algorithms. First, the 
contribution of parallactic angle and ionosphere were calibrated out. The visibility 
phases in the different sub-bands were aligned by fringe-fitting\footnote{See 
Schwab \& Cotton (\cite{Schwab}) for a detailed description of the global fringe-fitting 
(GFF) algorithm.} a scan of a strong source (3C345, in our case), following a procedure 
commonly known as ``manual phasecal''. Then, the multiband delays and rates of the resulting 
visibilities of all calibrators were fringe-fitted. The resulting gains (also corrected for 
the tropospheric effects with the AIPS task DELZN) were then interpolated into the scans of 
SN\,2011dh. The a priori amplitude calibration was based on the gain curves of each antenna 
and the system temperatures measured at each station. 

After this calibration, the behavior of the visibility amplitudes was checked as a function 
of time (for the different baselines) and as a function of baseline length (for similar directions in 
Fourier space), in a search for any parts of the experiment with bad amplitude calibration. Then, 
edition was applied to the obvious amplitude outliers. An image of the phase-calibrator 
source, J1332+4722, was generated by applying phase self-calibration (every 1--2 minutes) 
and a later amplitude self-calibration (one solution per antenna) until convergence in 
the gains and the model image was achieved. The recovered flux density was 
compared to the flux-density measurements performed with the Effelsberg radio telescope. We obtained 
compatible values of the flux density between Effelsberg, (307 $\pm$ 93)\,mJy, and the VLBI image, 
(231 $\pm$ 10)\,mJy. 

The antenna-based amplitude factors found in the imaging and self-calibration of the J1332+4722 
visibilities were then applied to the SN2011dh visibilities. Finally, an image of SN\,2011dh was 
generated from the calibrated visibilities, as described in the following section.

\section{Results and discussion}
\label{III}

\subsection{VLBI detection of SN2011dh}

\begin{figure}
\centering
\includegraphics[width=8.5cm,angle=270]{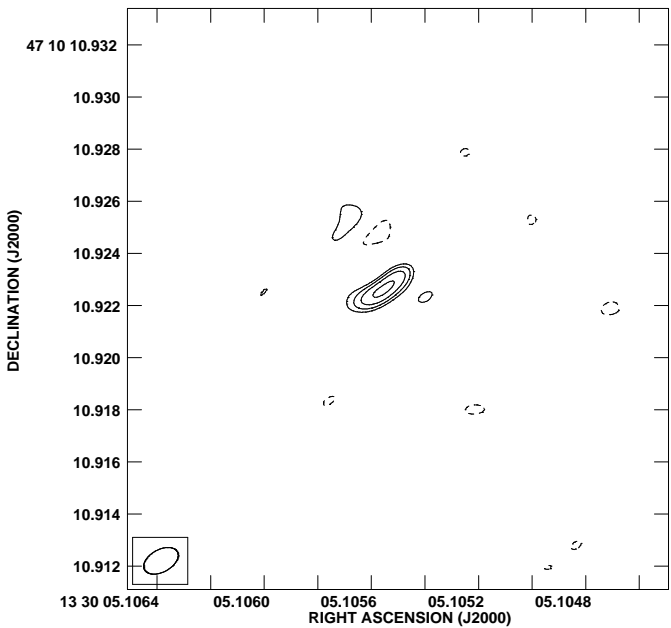}
\caption{Image of SN\,2011dh, phase referenced to J1332+4722. The full width 
at half maximum (FWHM) of the convolving beam is shown at the lower-left corner. The beamsize 
(at the FWHM) is 1.4$\times$0.9\,mas with PA = 61\,deg. (north to west). The contour levels are at
$-30$, 30, 45, 65, and 90\% of the peak flux density (2.2\,mJy\,beam$^{-1}$).
}
\label{fig1}
\end{figure}

After the calibration described in the previous section, an image of the supernova was 
synthesized by Fourier inversion from the space of visibilities into the sky plane. Natural 
weighting was applied to the visibilities, in order to improve the sensitivity of the 
array\footnote{With this scheme, the weight of each pixel in the Fast Fourier Transform (FFT) 
is inversely proportional to the scatter of the visibilities.}. 
We show the resulting image of the supernova in Fig. \ref{fig1}. Due to the limited dynamic 
range achieved, no further self-calibration was applied to the data to avoid the eventual 
generation of a spurious contribution to the flux density (see, e.g., Mart\'i-Vidal \& 
Marcaide \cite{MartiSpurious}). 

There is a clear detection of a compact 
source with a dynamic range of $\sim$10 and flux density of $2.5\pm0.5$\,mJy, which we identify 
as SN\,2011dh. 
According to the phase-reference calibration with respect to J1332+4722, 
the J2000.0 coordinates of the image peak
are $\alpha = 13^{\textrm{h}}\,30^{\textrm{m}}\,5.105559^{\textrm{s}}$ ($\pm 0.000007^{\textrm{s}}$) 
and $\delta = +47^{\circ}\,10'\,10.9226''$ ($\pm 0.0001''$). The uncertainties above contain 
the contribution from the error in the estimate of the calibrator position and the error inherent 
to the phase-referencing calibration, as estimated from Pradel et al. (\cite{Pradel2006}).

The source is very compact, and we do not see any clear
hint of structure on the scale of the synthesized angular resolution. Indeed, the fit of a 
uniform-disk model to the visibilities results in a size compatible with zero (upper limit of 
0.45\,mas, 1$\sigma$ cuttoff, for the disk radius). We notice that the maximum source size that is
compatible with our observations is similar to the size of our synthesized beam (as is indeed 
expected in observations of compact sources with a low signal-to-noise ratio, SNR). 
We therefore conclude that the angular size of SN\,2011dh is well below our resolution limit at this 
epoch of observations. This is an expected 
result, since the expansion velocity in the shocks of these types of supernovae is typically a 
few 10\,000\,km\,s$^{-1}$, at most. (The early ejecta velocity of SN\,2001dh from the 
$H\alpha$ is, indeed, 
17600\,km\,s$^{-1}$, Silverman et al. \cite{Silverman2011}.) Assuming a distance to M\,51 of 
7--8\,Mpc 
(e.g., Tak\'ats \& 
Vink\'o \cite{Takats2006}), a size similar to our beamwidth at this epoch would have implied 
a superluminal expansion. However, it is intriguing that, even though the emission comes from a very 
compact region at this epoch, we are unable to recover the total flux density measured
with the EVLA by Soderberg et al. (\cite{Soderberg2011}). The lack of flux density in our VLBI 
observations may come either from a contribution from an extended source in the EVLA observations
or from calibration biases in the VLBI (and/or EVLA) data, as discussed in Sect. \ref{Flux}.

\subsection{VLBI vs. EVLA flux density}
\label{Flux}

A systematic difference between the single-dish flux density of a source and the total recovered flux density in a 
VLBI-synthesized image is quite common. This effect takes place at all VLBI observing frequencies 
(with varying strength, but typically ranging from 10 to 20\% at most), and is due to either a limited 
coherence in the correlation or to extended components in the sources that are resolved out in the VLBI fringes. 
However, it must be noticed that the flux density of SN\,2011dh reported in Soderberg et al. 
(\cite{Soderberg2011}) implies a flux density of about 5\,mJy at 22\,GHz at the epoch of our observations 
(see their Fig. 3). This value is a factor 2 above the recovered flux density in our observations.
The ratio of single-dish-to-VLBI flux density of SN\,2011dh should be similar to that of J1332+4722, since the 
instrumental effects in the VLBI observations should be very 
similar in both sources. As a result, the Effelsberg-to-VLBI flux-density ratio of the calibrator, J1332+4722,
should also be similar to the EVLA-to-VLBI flux-density ratio for SN2011dh. This similarity 
should allow us to estimate the amount of lost flux density in the VLBI image of SN2011dh, and to compare it 
to the extra recovered EVLA flux density. 
However, the uncertainty in the flux-density measurement at the Effelsberg radio 
telescope is so high that such a comparison gives no statistical significance. The flux density of J1332+4722 
recovered with the Effelsberg radiotelescope is $1.33\pm0.45$ times higher than that recovered in the VLBI image.
Therefore, the relative amount of flux-density loss in the VLBI image of SN\,2011dh seems to be slightly larger 
than (although still compatible with) the one found in the image of the phase-reference calibrator. Nevertheless, 
the level of significance for this statement is very low.

It must also be noticed that Bietenholz et al. (\cite{Bieten2011}) report a VLBI (VLBA + 
GBT + Effelsberg) peak intensity for SN\,2011dh of only 0.63\,mJy at 22\,GHz (on day 83 after the optical discovery) 
using the same phase calibrator (J1332+4722). This peak intensity is also lower than what would be expected 
from the extrapolation of the model reported in Soderberg et al. (\cite{Soderberg2011}).

In addition to the well-known loss of flux density in VLBI images, caused mainly by the resolution of extended emission, 
there is an additional loss of flux density from the effect of atmospheric turbulence in 
phase-referenced observations (Mart\'i-Vidal et al. \cite{Marti2010coh}). 
To estimate the atmospheric contribution to the flux-density loss in our VLBI observations, we measured 
the total flux density in the image of B1333+459, phase-referenced to 
J1332+4722, and compared it to the total flux density in the image obtained from the self-calibrated visibilities of 
B1333+459. The observed flux-density loss in B1333+459 due to phase referencing is $\sim$40\%, which is within a factor 
2 of the loss predicted in Mart\'i-Vidal et al. (\cite{Marti2010coh}) for the VLBA ($\sim$25\%, as computed from their 
Eq. 4). These results indicate a similar performance of the EVN and the VLBA in phase-referencing observations, 
even using a very small subset of the EVN in our observations (only four stations had useful detections; see Sect. \ref{II}) 
and a higher observing frequency. If Eq. 4 in Mart\'i-Vidal et al. (\cite{Marti2010coh}) is used to estimate 
the flux-density loss of SN\,2011dh due to phase referencing, the result is only $\sim$2\% (using the constants reported 
in that publication, which are based on VLBA observations) or $\sim$3.5\% (if we calibrate Eq. 4
using our estimated flux-density loss of B1333+459, phase-referenced to J1332+4722). In any case, the expected 
flux-density loss of SN\,2011dh due to phase referencing (based on Eq. 4 of Mart\'i-Vidal et al. \cite{Marti2010coh}) 
is very small, given the small separation from its phase calibrator (about 3.4 times smaller than the distance 
between B1333+459 and J1332+4722).
 
In case that the missing flux density in our SN\,2011dh VLBI observations (and in those reported by 
Bietenholz et al. \cite{Bieten2011}) was not completely due to 
instrumental limitations (no robust conclusion can be extracted only from our data), such a loss might also be 
related to a contribution of extended emission (e.g., from the background galaxy, from the continuum of the whole region, or 
even from the supernova environtment) in the EVLA flux densities reported in Soderberg et al. (\cite{Soderberg2011}). 
We note, though, that such a contribution from an extended component in the supernova would imply that there is 
extended emission as strong as what comes from the (still compact) expanding shock, and this would thus conflict 
with the model reported in Soderberg et al. (\cite{Soderberg2011}), which assumes that all the emission 
detected with the EVLA comes from interaction of the expanding shock (with an expansion velocity of 
$\sim$30\,000\,km\,s$^{-1}$).

\section{Summary}
\label{IV}

We report on the VLBI detection of SN\,2011dh at 22\,GHz using a subset of the EVN array. The observations
took place 14 days after the discovery of the supernova. Therefore, this is the VLBI image of the 
youngest radio-loud supernova. The source is very compact, with a size compatible with zero and upper 
bound of 0.45\,mas for the radius of a uniform-disk model fitted to the visibilities). 

We provide revised coordinates for the supernova with milli-arcsecond resolution and linked to the ICRF. 
The recovered flux density is a factor $\sim$2 below the flux density reported in Soderberg et al. 
(\cite{Soderberg2011}), at the same frequency and day as our observations. Such a difference may 
indicate a contribution by extended emission in the EVLA flux densities or calibration problems 
in the VLBI and/or the EVLA observations. Further VLBI observations of this supernova will be decisive 
in helping resolve this conflict.

\begin{acknowledgements}

The European VLBI Network is a joint facility of European, Chinese, South African, and 
other radio astronomy institutes funded by their national research councils. We acknowledge
the EVN chair and related staff for the swift answer and scheduling of the VLBI observations.
The single-dish flux densities reported are based on observations with the 100-m telescope of the MPIfR.
This research has been partially
supported by projects AYA2009-13036-C02-01 and AYA2009-13036-C02-02 of the MICINN and by grant 
PROMETEO 104/2009 of the Generalitat Valenciana. E.R. was partially supported
by the COST action MP0905 ``Black Holes in a Violent Universe''.

\end{acknowledgements}

\end{document}